\documentclass[12pt]{article}
\usepackage{amsmath,amsfonts,mathtools,amssymb,ascmac}
\usepackage{bm} 
\usepackage{amsthm}
\usepackage[linesnumbered,ruled,vlined]{algorithm2e}

\usepackage{listings}
\usepackage{xcolor}

\theoremstyle{definition}
\newtheorem{definition}{Definition}
\usepackage[hidelinks]{hyperref}   
\usepackage{setspace} 

\usepackage{booktabs}
\usepackage{array}
\usepackage{comment}
\usepackage{pifont}

\usepackage{graphicx}
\usepackage{here}
\usepackage{subcaption}
\usepackage{url}
\usepackage[font=small,labelfont=bf]{caption}

\usepackage[title]{appendix}
\usepackage[
  top=25mm,
  bottom=25mm,
  left=25mm,
  right=25mm
]{geometry}

\usepackage{natbib}

\newcommand{\thresh}{\tau}
\newcommand{\nlambda}{n_{\lambda}}
\newcommand{\rpackage}[1]{\texttt{#1}}

\title{Data-driven configuration tuning of \texttt{glmnet} for balancing accuracy and computational efficiency}
\author{
    Shuhei Muroya$^{1}$\thanks{Email: muroya.shuhei.697@s.kyushu-u.ac.jp} and Kei Hirose$^{2}$\thanks{Email: hirose@imi.kyushu-u.ac.jp}\\
    \small $^{1}$Joint Graduate School of Mathematics for Innovation, Kyushu University, Fukuoka, Japan\\
    \small $^{2}$Institute of Mathematics for Industry, Kyushu University, Fukuoka, Japan
}
\date{}

\begin{document}

\maketitle

\begin{abstract}
The \rpackage{glmnet} package in $\tt R$ is widely used for lasso estimation because of its computational efficiency.
Despite its popularity, \rpackage{glmnet} occasionally yields solutions that deviate substantially from the true ones because of the inappropriate default configuration of the algorithm.
The accuracy of the obtained solutions can be improved by appropriately tuning the configuration.
However, such improvements typically increase computational time, resulting in a tradeoff between accuracy and computational efficiency.  
Therefore, a systematic approach is required to determine the appropriate configuration.
To address this need, we propose a unified data-driven framework specifically designed to optimize the configuration by balancing solution path accuracy and computational cost. 
Specifically, we generate a large-scale training dataset by measuring the accuracy and computation time of \rpackage{glmnet}.
Using this dataset, we construct neural networks to predict accuracy and computation time from data characteristics and configuration.
For a new dataset, the proposed framework uses the trained networks to explore the configuration space and derive a Pareto front that represents the tradeoff between accuracy and computational cost.
This front enables automatic selection of the configuration that maximizes accuracy under a user-specified time constraint.
The proposed method is implemented in the $\tt R$ package \rpackage{glmnetconf}, available at \url{https://github.com/Shuhei-Muroya/glmnetconf.git}.
\end{abstract}

\small \textit{Keywords}: lasso, \texttt{glmnet}, hyperparameter optimization, computational efficiency

\section{Introduction}\label{intro}
The least absolute shrinkage and selection operator (lasso; \citealp{tibshiraniregression1996}) is a popular method for regression that uses an $\ell_1$ penalty to obtain sparse regression coefficients.
It can handle high-dimensional data, where the number of predictors exceeds the number of observations, and it provides interpretable results.
Owing to these features, the lasso is widely applied across various fields,  
such as signal processing \citep{candes2008introduction}, genomics \citep{bocelstad2007maicroarraylasso}, and astronomy \citep{Lu2015astrolasso}.

Let $N$ be the number of observations and $p$ be the number of predictors.
Let $\bm{X} \in \mathbb{R}^{N\times p}$ be the design matrix with rows $\boldsymbol{x}_i \in \mathbb{R}^p$ for $i=1,\dots,N$, and let $\bm{y} \in \mathbb{R}^{N}$ be the response vector.
We assume that the explanatory variables are standardized and the response vector is centered.
Under these assumptions, the lasso estimates the coefficient vector $\bm{\beta} \in \mathbb{R}^p$ by solving
\begin{equation}\label{eq:lasso}
    \underset{\boldsymbol{\beta}}{\text{minimize}}\quad
    \frac{1}{2N}
    \| \boldsymbol{y}-\bm{X}\boldsymbol{\beta}\|_2 ^2 
    +\lambda \| \boldsymbol{\beta} \|_1,
\end{equation}
where $\lambda>0$ is a regularization parameter, and $\|\cdot\|_2$ and $\|\cdot\|_1$ denote the $\ell_2$- and $\ell_1$-norms, respectively.
The lasso solution does not generally have a closed-form expression because of the nondifferentiability of the $\ell_1$ norm. Various algorithms have been proposed to solve the lasso problem \citep{Fu1998shooting,Osborne2000homotopy,efronleast2004,daubechies2004ista,amir2009fista,friedmanregularization2010,boyd2011admm}. 
In particular, the coordinate descent algorithm \citep{Fu1998shooting, friedmanregularization2010}
and the least angle regression (LARS) algorithm \citep{efronleast2004} have been widely used.
The coordinate descent algorithm provides a fast approximate solution by iteratively updating each coefficient. 
In contrast, LARS yields the exact entire solution path for the lasso problem but at a higher computational cost.
The coordinate descent algorithm and the LARS algorithm are implemented in the $\tt R$ package \rpackage{glmnet} and \rpackage{lars}, respectively.
The \rpackage{glmnet} package is widely used owing to its computational efficiency.
It was downloaded over 1.3 million times in 2024, exceeding 10 times the downloads of \rpackage{lars}, according to the Comprehensive R Archive Network (CRAN) download logs provided by \rpackage{cranlogs} \citep{cranlogs2019}.

\begin{figure}[t]
    \centering
    \begin{subfigure}{0.3\textwidth}
        \centering
        \includegraphics[width=\textwidth]{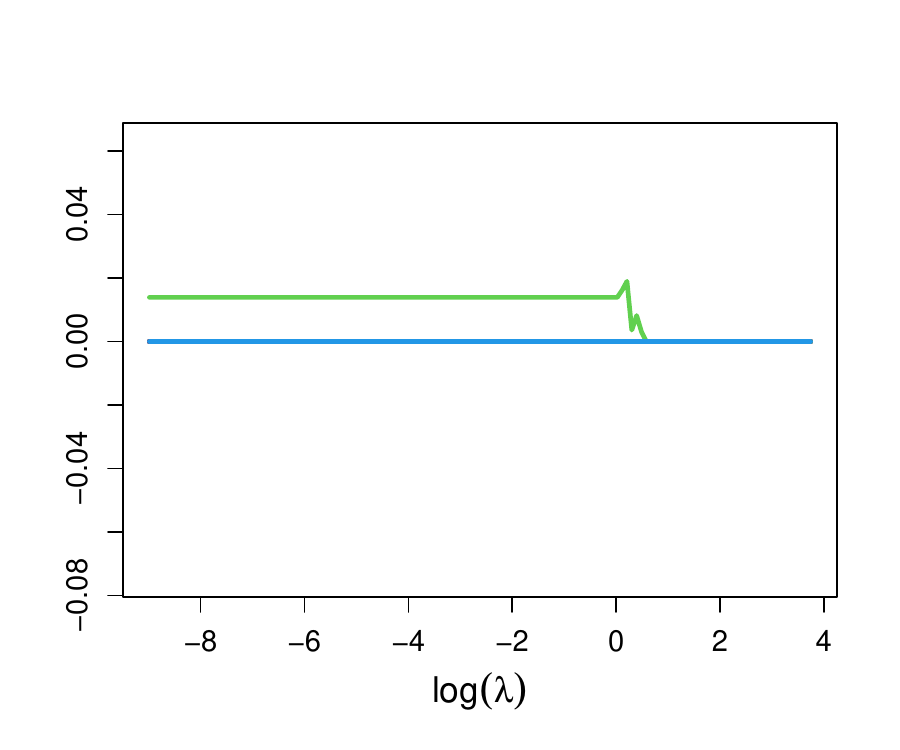}
        \caption{\rpackage{glmnet} (default)}
        \label{fig:sub1}
    \end{subfigure}
    \hfill
     \begin{subfigure}{0.3\textwidth}
        \centering
        \includegraphics[width=\textwidth]{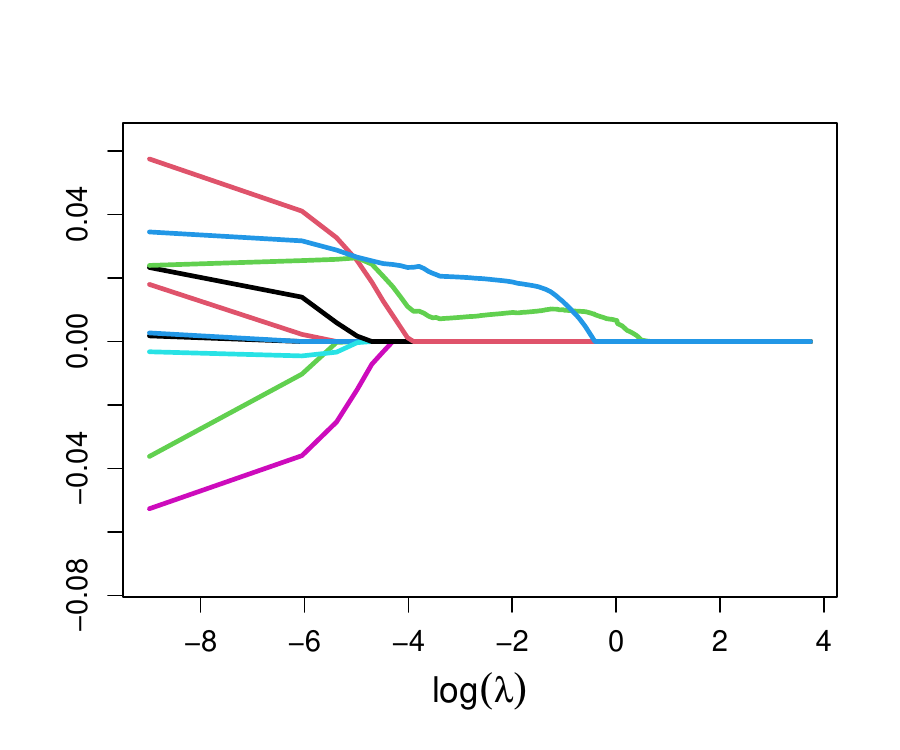}
        \caption{\rpackage{glmnet} (manual)}
        \label{fig:sub2}
    \end{subfigure}
    \hfill
    \begin{subfigure}{0.3\textwidth}
        \centering
        \includegraphics[width=\textwidth]{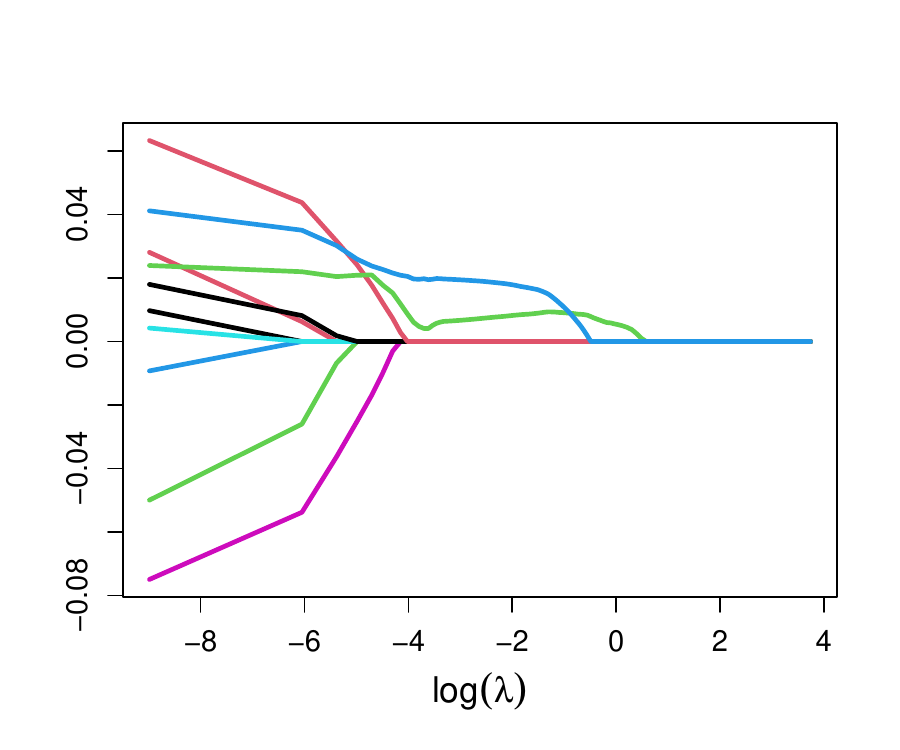}
        \caption{LARS}
        \label{fig:sub3}
    \end{subfigure}
    
    \caption{Solution paths for the same dataset obtained using different methods: (a) \rpackage{glmnet} (default); (b) \rpackage{glmnet} (manual); (c) LARS.
    The experimental setting is identical to that in Section \ref{sec: numericalexp}, with $N=1500, p=800, \rho=0.5$. 
    For clarity, we display the solution path for only the first 10 coefficients to avoid visual congestion.
    The \rpackage{glmnet} (default) denotes the estimator by \rpackage{glmnet} using the default configuration,
    whereas \rpackage{glmnet} (manual) denotes the estimator by \rpackage{glmnet} whose configuration is manually optimized by the authors.
    LARS denotes the estimator by the LARS algorithm, which provides the exact solution path and thus serves as a reference (ground truth).
    Notably, the results of \rpackage{glmnet} (manual) are close to that of LARS.
    }
    \label{fig:solutionpath}
\end{figure}

However, our numerical experiments reveal that the \rpackage{glmnet} solution path can deviate significantly from the exact solution path for correlated high-dimensional data.
These discrepancies may arise from the default settings in \rpackage{glmnet}.
In particular, the convergence threshold and the specification of the $\lambda$ sequence play critical roles.
Hereafter, we refer to these settings as the configuration of \rpackage{glmnet}.
To illustrate how the configuration affects the results, Figure~\ref{fig:solutionpath} compares the solution path of the first 10 coefficients obtained from three methods for a given dataset: \rpackage{glmnet} (default), \rpackage{glmnet} (manual) and LARS.
The \rpackage{glmnet} (default) and \rpackage{glmnet} (manual) denote the estimators obtained using the default and manually tuned configuration, respectively.
The label LARS corresponds to the exact solution path computed by the \rpackage{lars} package.
As shown in the figure, the solution path of \rpackage{glmnet} (default) is substantially different from that of LARS,
whereas the path of \rpackage{glmnet} (manual) is closer to that of LARS.
This result implies that appropriate tuning of the configuration is crucial for obtaining an accurate solution path.

In practice, users often rely on the default configuration without realizing its critical impact. This is partly because \rpackage{glmnet} returns results without warning, even when the default configuration is inappropriate for the given dataset. 
Furthermore, manual tuning is rarely performed as it requires expert knowledge of the underlying algorithm.
Notably, improving accuracy typically increases computational time, resulting in a tradeoff between accuracy and computational efficiency.
Although an appropriate configuration should ideally be determined for each dataset to balance this tradeoff, a systematic approach for such tuning has not yet been established.

Therefore, we propose a data-driven framework that automatically determines an appropriate configuration of \rpackage{glmnet} based on the characteristics of the dataset.
Specifically, we generate a large-scale training dataset by measuring the accuracy and computation time of \rpackage{glmnet}.
This dataset is used to train a neural network that learns the relationship among data characteristics, configuration, and their corresponding performance.  
Once trained, the neural network can predict accuracy and computation time for new datasets and configurations.
Based on these predictions, the Pareto front is derived to capture the tradeoff between accuracy and computation time.
From this front, our proposed framework automatically selects the configuration that achieves the highest possible accuracy, while ensuring that it does not exceed the user-specified computation time.

A key feature of our framework is its ability to explicitly manage the tradeoff between accuracy and computation time.
This capability enables users to perform configuration tuning that explicitly accounts for computational costs.
Furthermore, the tuning process is fast, thereby maintaining the total runtime shorter than that of LARS.

The organization of the paper is as follows.
In Section~\ref{sec:defaut_config}, we present the background and motivation for configuration tuning.
In Section~\ref{sec:proposed_method}, we explain our proposed method and demonstrate how to tune the configuration of the \rpackage{glmnet} function based on the dataset.
Section \ref{sec: numericalexp} evaluates the performance of our proposed method
through numerical simulations and the application to compressed sensing, respectively.

\section{Algorithmic details and default configuration in \rpackage{glmnet}}\label{sec:defaut_config}
This section reviews the computational details of the coordinate descent algorithm and the default configuration of \rpackage{glmnet}.
We specifically discuss why this default configuration can result in inappropriate solutions.
In addition, we briefly describe the LARS algorithm as a reference for the exact solution path.
\subsection{Coordinate descent algorithm}\label{sec:coordinatedescent}
The \texttt{R} package \rpackage{glmnet} implements the coordinate descent algorithm to solve the lasso problem efficiently \citep{friedmanregularization2010}.
For a given value of $\lambda$, 
the coordinate descent algorithm computes an approximate solution through an iterative procedure.
The entire solution path is obtained by repeatedly applying the algorithm over a sequence of $\lambda$ values.
Linear interpolation of these solutions yields an approximate solution path.

The coordinate descent algorithm optimizes one coefficient at a time 
while holding the others fixed, and cycles through all coefficients until convergence.
For a fixed value of the regularization parameter $\lambda$, 
the \rpackage{glmnet} package minimizes the objective function in \eqref{eq:lasso}
by iteratively updating each coefficient using the coordinate descent algorithm.
At iteration $t$, the update for the $j$-th coefficient $\beta_j^{t+1}$ is expressed by
\begin{equation*}
    \beta_j^{t+1}=
    S_{\lambda}\!\left(
        \frac{1}{N}\, \bm{X}_j^{\top} \bm{r}^{(j)}
    \right),
\end{equation*}
where $\bm{X}_j$ denotes the $j$-th column of $\bm{X}$, $S_{\lambda}(z) = \mathrm{sign}(z) \, (|z| - \lambda)_+$ is the soft-thresholding operator, and $\bm{r}^{(j)}$ represents the partial residual vector with elements $r_{i}^{(j)} = y_i - \sum_{k \neq j} x_{ik} \beta_k^{t}(\lambda)$.
This procedure cyclically updates all coefficients until convergence.

\subsection{Configuration details: Convergence threshold and the sequence of $\lambda$}\label{hypara}
This subsection examines the roles and default settings of two key components: the convergence threshold and the sequence of $\lambda$ values.
They are commonly used in both the \texttt{glmnet()} and \texttt{cv.glmnet()} functions of the \rpackage{glmnet} package.
Here, the function \texttt{glmnet()} computes a solution path on a grid of $\lambda$ values, 
whereas \texttt{cv.glmnet()} performs cross-validation to select the optimal $\lambda$ from this path.

\paragraph{Convergence threshold.}
The convergence threshold, denoted by $\thresh$, determines the stopping criterion for the coordinate descent algorithm.
Specifically, the iterative updates terminate when the improvement in the objective function falls below the product of $\thresh$ and the null deviance.
The smaller the $\thresh$, the stricter is the stopping condition. This can improve the accuracy of the solution but also increases computation time.
The default value of $\thresh$ is $10^{-7}$.  

\paragraph{$\lambda$ sequence.}
The sequence of $\lambda$ values defines the grid over which
the lasso solution path is computed, as discussed in the previous subsection.
Extending the range of $\lambda$ and refining the grid yields a more accurate solution path, but increases computation time.

The $\lambda$ sequence is determined by its range, defined by the maximum value $\lambda_{\max}$ and the minimum value $\lambda_{\min}$, and the number of grid points $\nlambda$.
In the \rpackage{glmnet} package, the default values for these parameters are specified as follows.
$\lambda_{\max}$ is defined as the smallest $\lambda$ for which all coefficients are zero, given by $\lambda_{\max} = \max_j \frac{1}{N}|\bm{X}_j^{\top}\bm{y}|$.
The default lower bound $\lambda_{\min}^{\mathrm{def}}$ is defined as:
\begin{equation*}
    \lambda_{\min}^{\mathrm{def}} = \begin{cases}
    10^{-2} \lambda_{\max} & \text{if } N < p, \\
    10^{-4} \lambda_{\max} & \text{if } N \ge p.
\end{cases}
\end{equation*}
Subsequently, the default sequence of $\lambda$ values is generated on a logarithmic scale 
from $\lambda_{\mathrm{max}}$ to $\lambda_\mathrm{min}^{\mathrm{def}}$ 
with $\nlambda^\mathrm{def} = 100$ points.

\subsection{Limitations of the default configuration}\label{whydefaultpoor}
We investigate the factors causing the default configuration of \rpackage{glmnet} to produce inaccurate results for highly correlated datasets.

\paragraph{Convergence threshold.}
Although the default threshold of $10^{-7}$ is computationally efficient, 
our numerical experiments suggest that a stricter threshold is often necessary to ensure accuracy.
This is required because of the following two main reasons:

First, high correlations among predictors lead to a flat lasso objective function.
In such cases, the coordinate descent algorithm moves in a zig-zag pattern with extremely small update steps.
Consequently, the improvement in the objective function at each step becomes extremely small, often causing the algorithm to terminate prematurely.
In addition, \citet{massias_2018_celer} noted that stopping rules based only on changes in the primal objective can lead to suboptimal solutions; they recommended monitoring the duality gap as a more rigorous criterion.
However, because we aim to improve \texttt{glmnet} without modifying its internal source code, we did not adopt the duality gap criterion.
Instead, we addressed this issue by using a significantly stricter threshold to improve the numerical precision.

Second, a stricter threshold is necessary to provide a more accurate initialization for the \textit{warm start} strategy.
As mentioned previously, the algorithm is repeatedly applied over a sequence of $\lambda$ values, denoted by 
$\lambda_1= \lambda_{\max} > \lambda_2 > \cdots > \lambda_{n_\lambda} =\lambda_{\min}$.
In this sequential process, the algorithm employs a \textit{warm start} strategy, where the solution obtained at the previous $\lambda$ is used to initialize the optimization for the current $\lambda$.
If the optimization at the previous step stops owing to a loose threshold, the resulting suboptimal solution provides an inaccurate starting point for the next step.
In a flat objective function, the solver may fail to move sufficiently away from this poor initialization, because the update steps are small and the stopping criterion is satisfied.
Consequently, the accumulation of such errors may cause the computed solution to deteriorate progressively.
Therefore, maintaining a tight convergence threshold is essential to prevent this error accumulation and to improve the reliability of the entire solution path.

\paragraph{$\lambda$ sequence.}
\begin{itemize}
    \item \textbf{Range of the sequence.} 
    The default sequence spans from $\lambda_{\max}$ to 
either $10^{-2} \lambda_{\max}$ or $10^{-4} \lambda_{\max}$, depending on whether $N<p$.
However, numerical experiments indicate that this range is occasionally extremely narrow 
to fully capture the behavior of the true solution path.
In particular, when compared with the exact path obtained by LARS, 
the default sequence often fails to explore the region of sufficiently small $\lambda$, 
where additional changes in the zero–-nonzero pattern can occur.
If these regions are omitted, the solution path computed by \texttt{glmnet} 
may miss important structural changes in the coefficients.
From the viewpoint of cross-validation, a narrow range restricts the diversity of candidate models.
Notably, the default sequence may fail to include the optimal $\lambda$, because the optimal $\lambda$ tends to be small when the correlation among predictors is high \citep{hebiri2013corlasso}.
Therefore, extending the range of the $\lambda$ sequence toward zero is essential to increase the probability that the optimal $\lambda$ is included in the candidate set.
    
    \item \textbf{Number of grid points.} 
The default number of grid points is $\nlambda^\mathrm{def} = 100$. However, this fixed number may be insufficient relative to the dimension $p$.
The LARS algorithm (Section \ref{subsec:lars}) implies that the active set of the lasso solution 
changes at least $\min\{N, p\}$ times along the path.
Thus, when $N$ and $p$ are larger than 100, the default grid cannot capture all changes in the 
true solution path, and linear interpolation between coarse grid points  
may degrade the accuracy of the approximated path.
From the perspective of cross-validation, a small number of $\lambda$ candidates 
means that the search space for selecting the $\lambda$ becomes extremely limited, 
which can result in suboptimal model selection.
\end{itemize}

The aforementioned discussions demonstrate that the default configuration, independent of the dataset, is insufficient to maintain numerical accuracy.
Although manual tuning of the configuration is possible, an automated approach tailored to the dataset is highly desirable in practice.
Therefore, we proposed a data-driven automated framework that determines the appropriate configuration to achieve accuracy comparable to LARS, while maintaining computational efficiency.

\subsection{LARS algorithm and solution path accuracy}\label{subsec:lars}
\citet{efronleast2004} proposed the LARS algorithm,
which provides an exact computation of the entire solution path of the lasso problem \eqref{eq:lasso}.
The algorithm begins at $\lambda = \infty$, where the lasso solution is trivially $\bm{0} \in \mathbb{R}^p$.
As $\lambda$ decreases, it computes a piecewise linear and continuous solution path.
Each knot along this path corresponds to a point where the active set 
$\mathcal{A} = \{ j : \beta_{j}(\lambda) \neq 0 \}$ changes.
At every iteration, the algorithm updates the direction of the coefficient path, ensuring 
that the Karush--Kuhn--Tucker (KKT) optimality conditions remain satisfied.
To determine this direction, the algorithm must compute 
the inverse of the Gram matrix $(\bm{X}_{\mathcal{A}}^{\top} \bm{X}_{\mathcal{A}})^{-1}$,
where $\bm{X}_{\mathcal{A}}$ denotes the submatrix of active predictors.
The active set $\mathcal{A}$ changes sequentially along the path; hence,
the LARS algorithm requires such matrix inversions to be performed at least $\min\{N, p\}$ times.
Consequently, the computational cost increases rapidly with the number of variables $p$.

In this study, we utilized LARS as a reference in three ways:
(i) the exact path served as the ground truth for evaluating approximation accuracy;
(ii) the exact number of knots was used to investigate the validity of the default $\lambda$ grid in \rpackage{glmnet}; and
(iii) the computation time provided an upper bound for efficiency comparisons.

\section{Proposed method}\label{sec:proposed_method}

\subsection{Overview of the proposed method}\label{optim_glmnethypera}
\begin{figure}
    \centering
    \includegraphics[width=\linewidth]{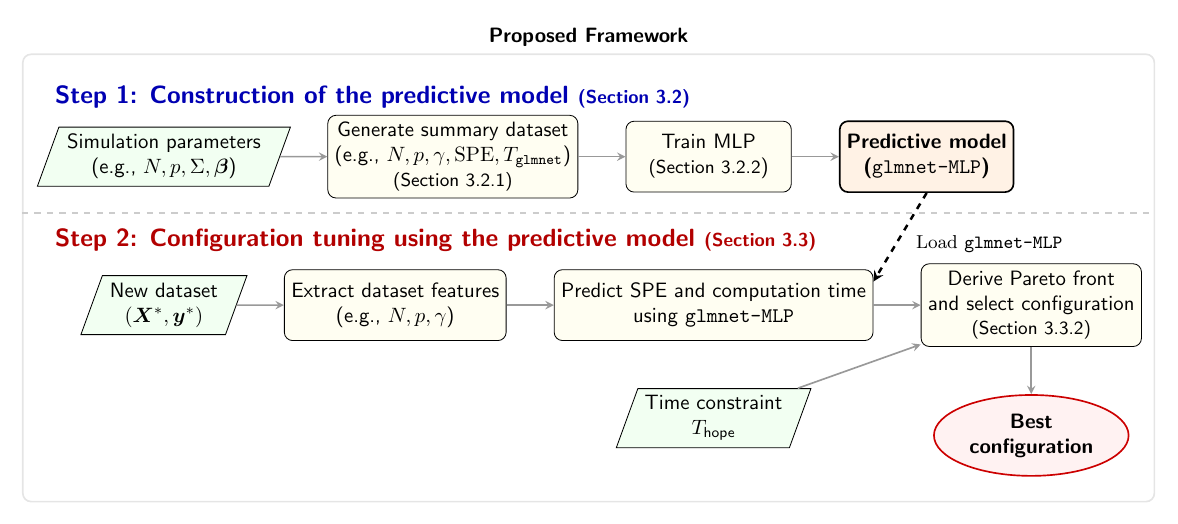}
    \caption{Overview of the proposed framework. The process comprises two phases: Step~1 constructs a predictive model using a summary dataset generated from simulation parameters. Step~2 utilizes this trained model to predict performance metrics for a target dataset, selecting the best configuration that satisfies the time constraint $T_{\text{hope}}$.}
    \label{fig: proposedframework}
\end{figure}

The proposed framework aims to automatically determine the appropriate configuration for a given dataset.
Specifically, it aims to maximize accuracy given a user-specified computation time, denoted as $T_{\text{hope}}$.
To this end, we focused on tuning two key parameters: the convergence threshold $\thresh$ and the sequence length $\nlambda$.
The detailed definition of $\nlambda$ is provided in Section \ref{step1_proposed}.
Figure~\ref{fig: proposedframework} illustrates the overall workflow of our proposed framework, which comprises two main steps:
\begin{itemize}
    \item \textbf{Step 1: Construction of the predictive model (Section~\ref{step1_proposed}).}
    The upper panel of Figure~\ref{fig: proposedframework} shows the preparatory stage.
    Starting from diverse simulation parameters, we generated a summary dataset to train a predictive model, which we refer to as \texttt{glmnet-MLP}. 
    This model learns the mapping between dataset characteristics (e.g., $N, p,\gamma$), configurations $(\thresh, \nlambda)$, and the resulting performance metrics, specifically the computation time and the Solution Path Error (SPE). 
    Sections~\ref{datagene} and \ref{training_strategy} provide the details of this process, including the formal definition of SPE, generation of the summary dataset, and training strategy.

    \item \textbf{Step 2: Configuration tuning using the predictive model (Section~\ref{step2_proposed}).}
    The lower panel of Figure~\ref{fig: proposedframework} presents the execution phase.
    Given a new dataset, the framework extracts its features and 
    utilizes the trained \texttt{glmnet-MLP} to predict performance.
    Finally, by deriving the Pareto front of the predicted SPE and computation time, the best configuration is automatically selected to maximize accuracy while satisfying the time constraint $T_\text{hope}$.
    The details of this tuning strategy are provided in Section~\ref{selectPareto}.
\end{itemize}

\subsection{Step 1: Construction of the predictive model}\label{step1_proposed}
In this step, we construct a predictive model using a multilayer perceptron (MLP) \citep{rumelhart_learning_1986}, referred to as \texttt{glmnet-MLP}.
The objective of this model is to predict the performance metrics, specifically the SPE and the computation time $T_{\mathtt{glmnet},\thresh, \nlambda}$.
Here, $T_{\mathtt{glmnet},\thresh, \nlambda}$ is defined as the total runtime, including the cross-validation process for selecting the optimal $\lambda$.

Regarding the dataset characteristics, we specifically included the sample size $N$, number of predictors $p$, and eigenvalue features of the covariance matrix.
The eigenvalues were included to capture the correlation structure among the predictors.

In addition to these data features, we incorporate the configuration parameters: the convergence threshold $\thresh$ and the length of the $\lambda$ sequence $\nlambda$.
One characteristic of our framework is the construction of the $\lambda$ sequence using $\nlambda$.
In contrast to the default configuration, we proposed a flexible construction where the sequence length is determined by $\nlambda$ ($\nlambda > \nlambda^\mathrm{def}$).
Specifically, we extend the default sequence by appending $(\nlambda - \nlambda^\mathrm{def})$ additional values evenly spaced between the default minimum ${\lambda_{\mathrm{min}}^\mathrm{def}}$ and $0$.
Through this parameterized construction, the complex problem of designing an appropriate $\lambda$ sequence was effectively reduced to determining a single optimal value for $\nlambda$.

\subsubsection{Construction of the summary dataset}\label{datagene}
To train the \texttt{glmnet-MLP}, we constructed a large-scale dataset, which we refer to as the summary dataset.
This dataset was created by generating an artificial dataset and recording the corresponding \rpackage{glmnet} performance.
Each sample in the summary dataset comprised the data characteristics ($N, p$, and eigenvalues), configuration ($\thresh, \nlambda$), and resulting performance metrics (SPE and computation time).
The detailed construction procedure is as follows:
\begin{enumerate}
    \item{Parameter setting and feature extraction:}
    Specify the simulation parameters: sample size $N$, number of predictors $p$, covariance matrix $\bm \Sigma$, true coefficients $\bm{\beta}$, and error variance $\sigma^2$.
    At this stage, we compute the eigenvalue features of $\bm \Sigma$. 
    Select the top and bottom five eigenvalues, denoted as $\gamma_k$ ($k = \pm 1, \dots, \pm 5$), where positive and negative indices correspond to the largest and smallest eigenvalues, respectively.

    \item{Data generation:}
    Using the specified parameters, we generated a synthetic dataset $(\bm{X}, \bm{y})$ according to
    \[
    \bm{x}_i \sim \mathcal{N}(\bm{0}, \bm \Sigma) \ (i = 1,\dots,N), \quad 
    \bm{\epsilon} \sim \mathcal{N}(\bm{0}, \sigma^2 \mathbf{I}), \quad 
    \bm{y} = \bm{X} \bm{\beta} + \bm{\epsilon},
    \]
    where $\mathcal{N}(\bm{\mu}, \bm{\Sigma})$ denotes the multivariate normal distribution with mean $\bm{\mu}$ and covariance matrix $\bm{\Sigma}$.

    \item{Performance evaluation:}
    We compute the lasso solutions using \rpackage{glmnet} under various configurations $(\thresh, \nlambda)$.
    For each configuration, we quantify the discrepancy between the approximate solution path and the exact path using SPE, defined as follows:
    \[
        \mathrm{SPE}_{\thresh, \nlambda} = \frac{1}{k}\sum_{i=1}^{k} \frac{1}{\sqrt{p}} \left\| \bm{\beta}^{\mathrm{true}}(\lambda_i) - \hat{\bm{\beta}}^{\mathrm{glmnet}}_{\thresh, \nlambda}(\lambda_i) \right\|_2,
    \]
    where $\{\lambda_i\}_{i=1}^k$ is a reference sequence of $k=20$ points logarithmically spaced from $\lambda_{\max}$ to $\lambda_{\mathrm{start}}=0.001$.
    Here, $\bm{\beta}^{\mathrm{true}}(\lambda_i)$ is the exact solution obtained via LARS, and $\hat{\bm{\beta}}^{\mathrm{glmnet}}_{\thresh, \nlambda}(\lambda_i)$ is the solution estimated by \rpackage{glmnet}.
     In addition, we recorded the computation time $T_{\mathtt{glmnet},\thresh, \nlambda}$.

    \item{Data aggregation:}
    We recorded the combination of the data characteristics, configuration, and performance metrics as a single data point:
    \[
    (N, p, \gamma_1, \dots, \gamma_{-1}, \thresh, \nlambda, \mathrm{SPE}_{\thresh,\nlambda}, T_{\mathtt{glmnet},\thresh, \nlambda}).
    \]

    \item{Iteration:}
    We repeated Steps 1--4 under various parameter settings.
    Consequently, this process yielded a total of $810{,}492$ samples, which constitute the summary dataset.
\end{enumerate}
Detailed specifications of the simulation parameters and the summary dataset are provided in Appendix \ref{appendix:parameter}.

\subsubsection{Training strategy and determination of network architecture}\label{training_strategy}
We trained the \texttt{glmnet-MLP} using the summary dataset.
The dataset was randomly split into training, validation, and test sets.
Prior to training, the target variables (SPE and computation time) were log-transformed and standardized to stabilize learning.
To obtain predictions on the original scale, we applied inverse transformations.

To determine the optimal network architecture (e.g., number of layers and units) and the learning rate, we employed Bayesian optimization.
We formulated the task as a black-box optimization problem to minimize the validation error and implemented it using the \texttt{Optuna} framework \citep{akiba2019optuna}.
Further details regarding the training protocol, the search space for hyperparameters, and the final network architecture are provided in Appendix \ref{app:training_details}.

\subsection{Step 2: Configuration tuning using the predictive model }\label{step2_proposed}

\subsubsection{Definition of Pareto front}\label{def:Pareto}
First, we introduce the concept of Pareto optimality for a multiobjective optimization problem.
We consider the problem of simultaneously minimizing a vector-valued objective function $\bm{f} : \mathcal{X} \to \mathbb{R}^M$:
\begin{equation}
    \underset{x\in \mathcal{X}}{\min} \ \bm{f}(x)=
    \underset{x\in \mathcal{X}}{\min} \left( f^{1}(x), \dots, f^{M}(x) \right). \label{multioptim}
\end{equation}
Generally, a unique solution that minimizes all objective functions simultaneously does not exist.
Instead, we seek Pareto optimal solutions, which represent optimal tradeoffs among the objectives.
\begin{definition}[Weak dominance]
    For $x,x'\in \mathcal{X}$, if $f^{m}(x)\leq f^{m}(x')\ \forall m=1,\cdots M$,
    we state that $\bm{f}(x)$ weakly dominates $\bm{f}(x')$.
\end{definition}
A Pareto optimal solution is defined as follows:
\begin{definition}[Pareto optimal solution and Pareto front]
    We state that \( x^* \in \mathcal{X} \) is a Pareto optimal solution if
    no \( x \in \mathcal{X} \) exists such that \( \bm{f}(x) \) weakly dominates \( \bm{f}(x^*) \)
    with \( \bm{f}(x) \neq \bm{f}(x^*) \).
    In addition, we define the Pareto front as the set of the objective values of Pareto optimal solutions.
    The Pareto front $\mathcal{P}^*$ is expressed by
    \begin{equation*}
        \mathcal{P}^*=\{\bm f(x^*)\mid x^*\in \mathcal{X}: \text{Pareto optimal solution}\}.
    \end{equation*}
\end{definition}

Theoretically, an infinite number of Pareto optimal solutions may exist.
Thus, we need to select the \textit{best} solution from the set of Pareto optimal solutions.

\subsubsection{Pareto front for optimizing configuration}\label{selectPareto}
In this section, we describe the procedure to tune the \rpackage{glmnet} configuration using the trained predictive model.
Our goal is to determine the optimal configuration $(\thresh^*, \nlambda^*)$ for a new dataset $(\bm{X}^*, \bm{y}^*)$ under a user-specified computation time constraint, denoted as $T_{\text{hope}}$.
The specific procedure is as follows:

\begin{enumerate}
    \item \textbf{Feature extraction:}
    We compute the data characteristics for the target dataset $(\bm{X}^*, \bm{y}^*)$. Specifically, we calculate the sample size $N$, dimension $p$, and eigenvalue statistics $\gamma_i \ (i = \pm1, \dots, \pm5)$ derived from the sample covariance matrix of $\bm{X}^*$. Notably, in contrast to the training phase (Step 1), where the eigenvalues were computed from the true covariance matrix $\bm{\Sigma}$, here they are derived from the sample covariance matrix of $\bm{X}^*$.

    \item \textbf{Model setup:}
    We fix these extracted features in \texttt{glmnet-MLP}. Consequently, the MLP functions as a mapping from configuration $(\thresh, \nlambda)$ to the predicted SPE and computation time. This mapping corresponds to the objective function $\bm{f}(x)$ in Eq.~(\ref{multioptim}).

    \item \textbf{Random sampling:}
     We randomly sample $K$ configurations $\{(\thresh^k, \nlambda^k)\}_{k=1}^K$ from the search space, where $\thresh$ is sampled from $[ 10^{-9}, 10^{-7}]$ on a log scale and $\nlambda$ from $[100, 2p]$.

    \item \textbf{Performance prediction:}
    We obtain the predictions $\{\mathrm{SPE}_{\thresh^k, \nlambda^k}, T_{\rpackage{glmnet},\thresh^k, \nlambda^k}\}_{k=1}^K$ by substituting the sampled configurations $\{(\thresh^k, \nlambda^k)\}_{k=1}^K$ into the mapping defined in the ``Model setup'' step.
    
    \item \textbf{Pareto front extraction:}
    We identify the discrete Pareto front $\widehat{\mathcal{P}}^*$ from the set of predicted outcomes $\{\mathrm{SPE}_{\thresh^k, \nlambda^k},\allowbreak T_{\rpackage{glmnet},\thresh^k, \nlambda^k}\}_{k=1}^K$.

    \item \textbf{Best configuration selection:}
    From the Pareto front $\widehat{\mathcal{P}}^*$, we select the optimal configuration $(\thresh^*, \nlambda^*)$ that minimizes the SPE subject to a user-specified computation time constraint $T_{\text{hope}}$.
    The index of the best configuration $k^*$ is determined by the following:
    \[
        k^* = \underset{k \in \{1,\dots,K\}}{\arg\min}\ \mathrm{SPE}_{\thresh^k, \nlambda^k}
        \quad \text{subject to} \quad
        \begin{cases}
            T_{\rpackage{glmnet},\thresh^k, \nlambda^k} < T_{\text{hope}}, \\
            \left(\mathrm{SPE}_{\thresh^k, \nlambda^k}, T_{\rpackage{glmnet},\thresh^k, \nlambda^k}\right) \in \widehat{\mathcal{P}}^*.
        \end{cases}
    \]
    
    Finally, the best configuration is given by $(\thresh^*, \nlambda^*) = (\thresh^{k^*}, \nlambda^{k^*})$.
\end{enumerate}

\begin{figure}[t]  
    \centering
    \includegraphics[width=1\textwidth]{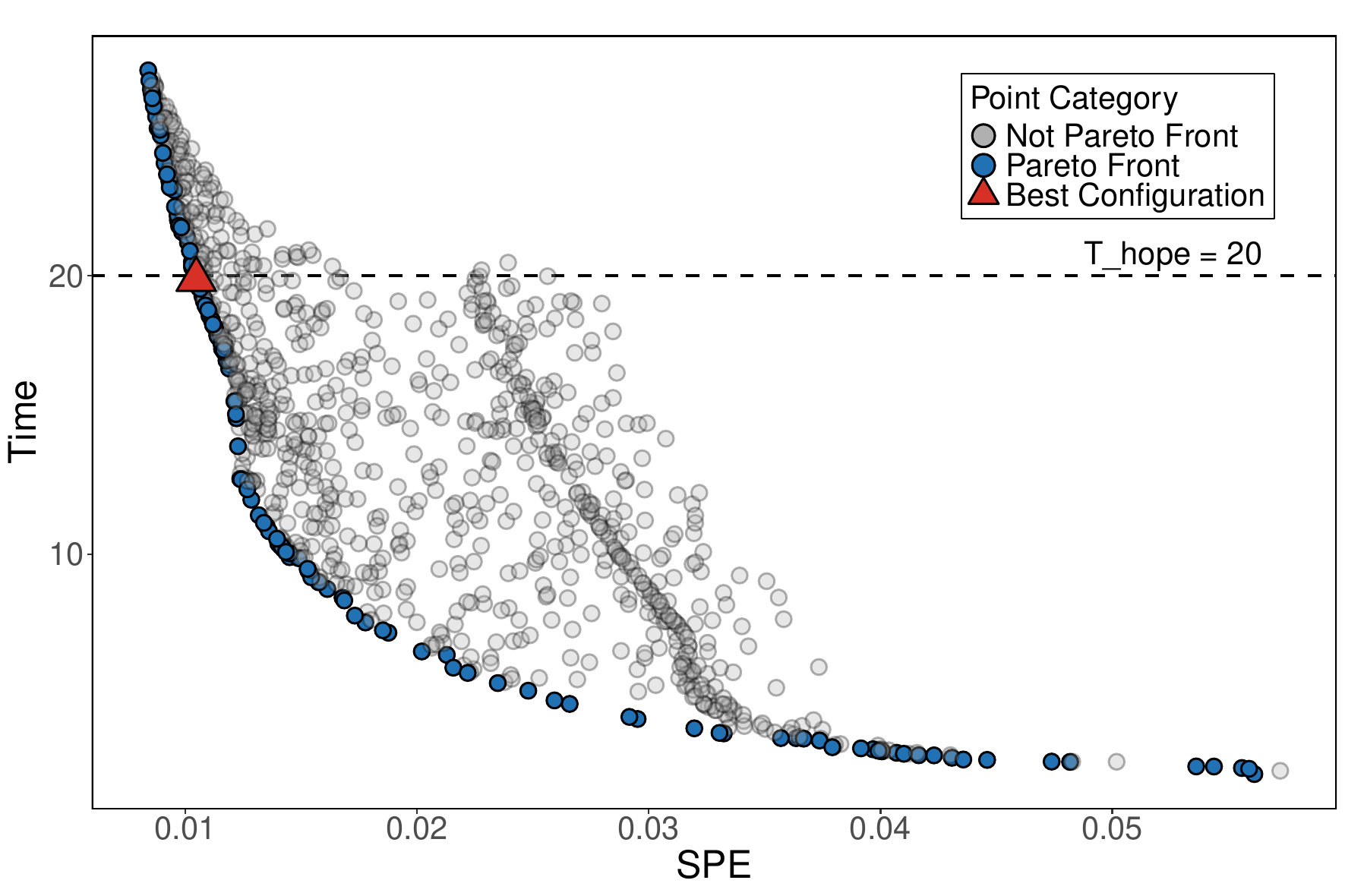}  
    \caption{Visualization of the Pareto front derived from the \texttt{glmnet-MLP} for the same dataset used in Figure~\ref{fig:solutionpath}. The horizontal and vertical axes represent the predicted SPE and computation time, respectively. The blue points represent the set of Pareto optimal solutions. From this set, the red triangle highlights the best configuration selected based on the user-specified time constraint ($T_{\text{hope}}=20~\mathrm{s}$), indicated by the horizontal dashed line.
    Under this constraint, the best configuration was identified as $(\thresh^*, \nlambda^*) = (1.159 \times 10^{-9}, 864)$.}
    \label{fig:ex1Paretofront}  
\end{figure}

By applying this tuning procedure to the same dataset used in Figure~\ref{fig:solutionpath}, we obtained the Pareto front shown in Figure~\ref{fig:ex1Paretofront}.
In this example, we set the time constraint to $T_{\text{hope}} = 20$ s.
This approach offers significant advantages in terms of both efficiency and interpretability.
First, the optimization process is extremely fast; for instance, computing the Pareto front for Figure~\ref{fig:ex1Paretofront} required only approximately 1 s.
The only computationally intensive step is the eigenvalue calculation.
Upon extraction, evaluating thousands of configurations via the neural network requires negligible time.
This efficiency meets the requirement discussed in Section~\ref{intro} to optimize the configuration as quickly as possible.
Second, the Pareto front provides visual clarity regarding the tradeoff between SPE and computation time.
This enables users to assess the cost of accuracy.
For example, in Figure~\ref{fig:ex1Paretofront}, we can observe a substantial difference in SPE between computation times of 20 s and 5 s.
Based on this visualization, users can make informed decisions, such as whether to relax or tighten the constraint $T_{\text{hope}}$ to achieve the desired balance.

We implemented the proposed framework as an \texttt{R} package named \rpackage{glmnetconf}.
This package provides the configuration tuning framework proposed in this study.
Furthermore, it incorporates a mechanism to select the appropriate package (i.e., \rpackage{glmnet} or \rpackage{lars}) considering computation time.
The details of this package selection and specific usage examples are provided in Appendix~\ref{glmnetconf}.

\section{Numerical experiments}\label{sec: numericalexp}
\subsection{Simulation}
In this section, we verify that our proposed method properly tunes the configuration $(\thresh, \nlambda)$ through numerical experiments.
The simulation dataset with $N$ observations and $p$ predictors was generated as follows:
\begin{align*}
    \bm{x}_i &\overset{\text{i.i.d.}}{\sim} \mathcal{N}(\bm{0},  (1-\rho) \ \bm{I}_p + \rho \,\bm{1}_p\bm{1}_p^\top), \quad \bm{X} = (\bm{x}_1, \dots, \bm{x}_N)^\top, \\ 
    \bm{\beta} &= \bm{P} ( \underbrace{1, \ldots, 1}_{\lfloor p/2\rfloor}, \underbrace{0, \ldots, 0}_{p-\lfloor p/2\rfloor} )^\top, \quad \bm{\varepsilon} \sim \mathcal{N}(\bm{0}, \bm{I}_N), \\ 
    \bm{y} &= \bm{X} \bm{\beta} + \bm{\varepsilon},
\end{align*}
where $\bm{P}$ is a random permutation matrix of size $p \times p$, and $\lfloor \cdot \rfloor$ denotes the floor function.
In this simulation, we compare the performance of the following three methods:
\begin{itemize}
    \item \rpackage{glmnet} (default): \rpackage{glmnet} with the default configuration.
    \item \rpackage{glmnet} (proposed): \rpackage{glmnet} with the configuration optimized by our proposed 
    method with $T_\text{hope}=20~\mathrm{s}$. 
    \item LARS: Serves as a reference to provide the exact solution path by the \rpackage{lars} package.
\end{itemize}

We conducted the experiments for all combinations of $N, p \in \{100, 500, 1000, 1500, 2000\}$ and $\rho \in \{0, \allowbreak 0.1, \allowbreak 0.3, \allowbreak 0.5, \allowbreak 0.7, \allowbreak 0.9\}$ over 100 simulation runs.
To evaluate the predictive performance, we employed the Root Mean Square Error (RMSE) computed on test datasets of 100 samples.
In addition, we measured the computation time for each method.
The regularization parameter $\lambda$ was selected via ten-fold cross-validation by \texttt{cv.glmnet()} and \texttt{cv.lars()}.
From the perspective of numerical stability, we specified \texttt{mode = "step"} in \texttt{cv.lars()} when $N = p$, while choosing \texttt{mode = "fraction"} otherwise.

\begin{figure}[t]
    \centering
    \includegraphics[width=\linewidth]{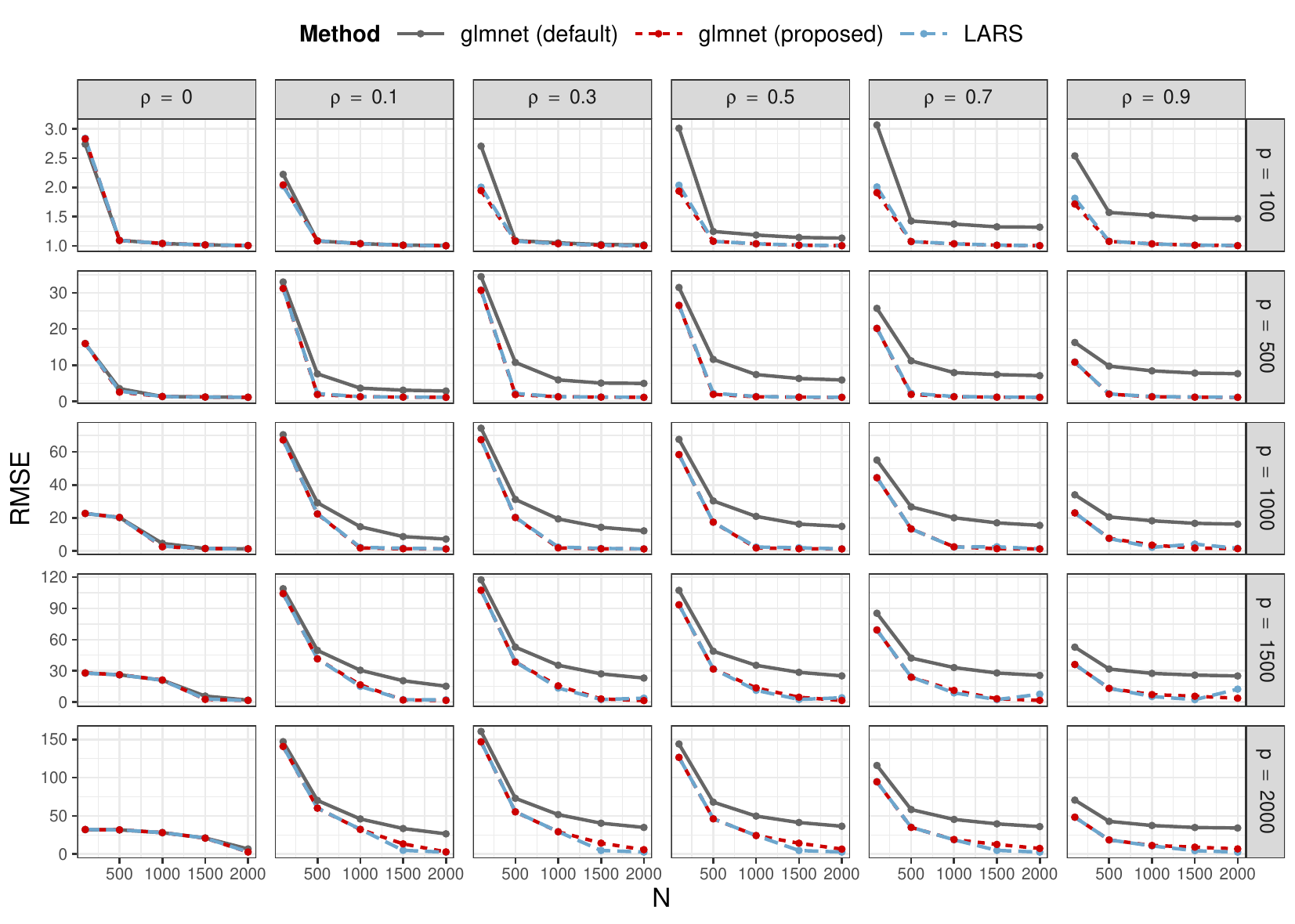}
    \caption{Comparison of prediction accuracy (RMSE) across different sample sizes $N$. 
    The plot compares \rpackage{glmnet} (default), \rpackage{glmnet} (proposed) tuned with $T_\text{hope}=20~\mathrm{s}$ and LARS as the exact reference.
    The results are averaged over 100 simulation runs. The panels correspond to different combinations of the number of predictors $p$ and the correlation among the predictors $\rho$.
    Notably, the \rpackage{glmnet} (proposed) consistently achieves accuracy comparable to the exact LARS solution across all settings.
    }
    \label{fig:simulation_rmse}
\end{figure}

Figure~\ref{fig:simulation_rmse} presents the results of the numerical experiment. 
In each panel, the vertical axis represents the average RMSE, and the horizontal axis represents the sample size $N$. 
The panels are organized by combinations of the number of predictors $p$ and the correlation $\rho$.
When $\rho = 0$, the test errors of all three methods are similar across all combinations of $N$ and $p$.
However, when $\rho>0$, the test error of \rpackage{glmnet} (default) is higher than that of LARS.
This result indicates that the default configuration is not appropriate for such correlated data. 
In contrast, \rpackage{glmnet} (proposed) achieveed performance comparable to that of LARS in most cases. Although slightly higher errors are observed when $p=2000$, this can be attributed to the imposed $T_\text{hope}$, reflecting the tradeoff between computational time and accuracy.

\begin{figure}[t]
    \centering
    \includegraphics[width=\linewidth]{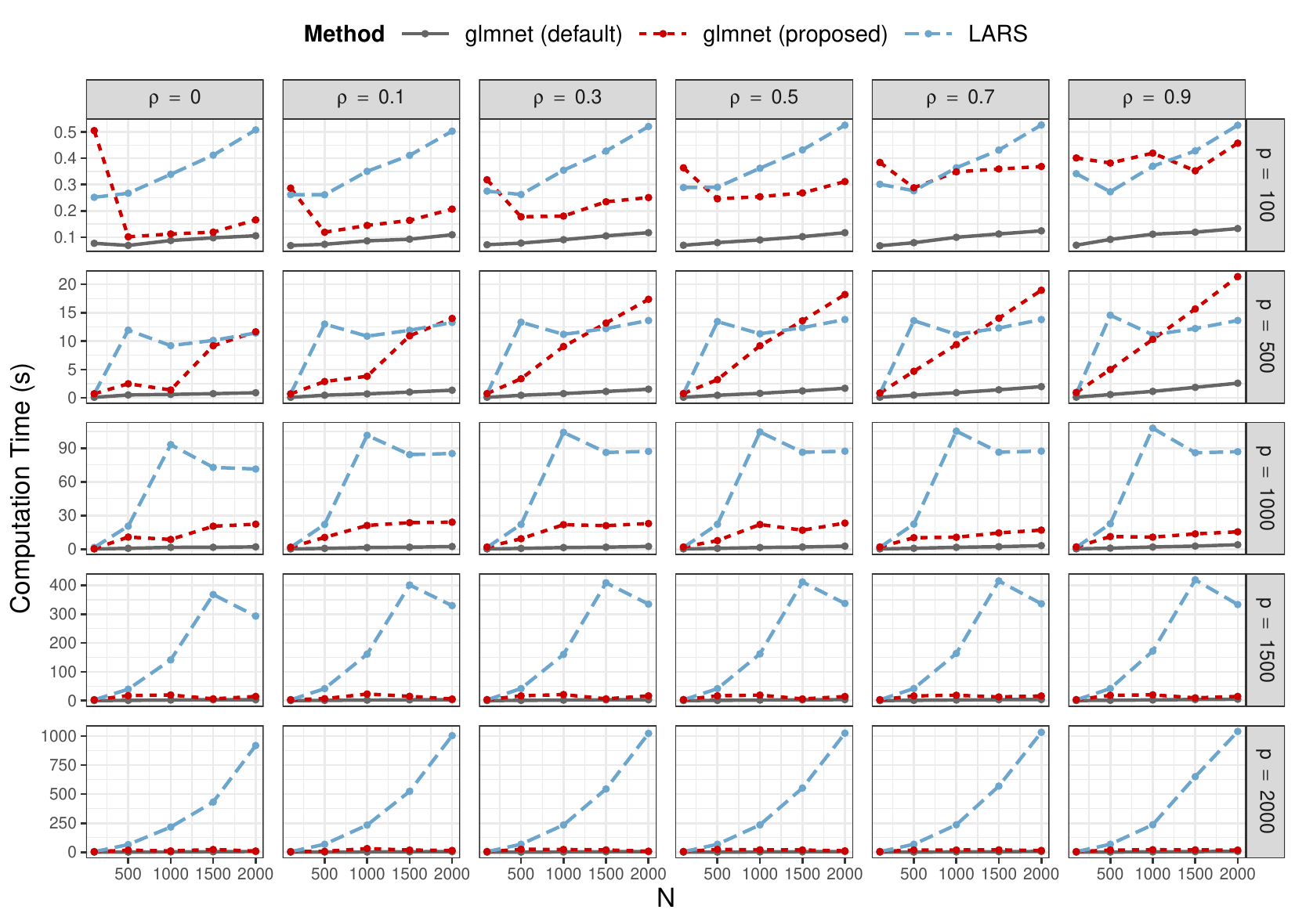}
    \caption{Comparison of computation time (seconds) across different sample sizes $N$.
    Similar to Figure~\ref{fig:simulation_rmse}, this plot compares \rpackage{glmnet} (default), \rpackage{glmnet} (proposed) tuned with $T_\text{hope}=20~\mathrm{s}$, and LARS.
    The results are averaged over 100 simulation runs. The panels correspond to different combinations of the number of predictors $p$ and the correlation among the predictors $\rho$.
    The proposed method is not only significantly faster than LARS but also satisfies $T_\text{hope}$ in most cases.}
    \label{fig:simulation_time}
\end{figure}

Figure~\ref{fig:simulation_time} reports the average computation time of the experiments using the same layout as Figure~\ref{fig:simulation_rmse}.
Among the three methods, LARS consistently required the longest computation time for $p \ge 1000$; its runtime increased drastically with larger $N$ and $p$. 
In contrast, \rpackage{glmnet} (proposed) was significantly faster in these settings, with runtimes consistently staying close to $T_{\text{hope}}$.
Despite this speed advantage, Figure~\ref{fig:simulation_rmse} confirms that their predictive accuracy remains comparable.
Overall, these results demonstrate that \rpackage{glmnet} (proposed) achieves accuracy comparable to that of LARS while significantly reducing computational time. 
This suggests that our proposed method successfully selects the appropriate configuration for \rpackage{glmnet} adaptively based on the dataset.
Notably, for $p \le 500$, \rpackage{glmnet} (proposed) occasionally exhibited slightly longer computation times than LARS. 
This behavior is attributable to the setting of $T_{\text{hope}}$. 
For small-scale problems, the computational budget is relatively generous, enabling \rpackage{glmnet} (proposed) to utilize the available time to maximize accuracy.

\subsection{Application to compressed sensing}\label{sec: compsense}
Compressed sensing \citep{candes2008introduction} is a signal processing technique that reconstructs a signal from a compressed representation obtained via a random projection matrix.
In this section, we apply our proposed framework to solve the lasso problem arising in compressed sensing.
We compare the reconstruction accuracy and computation time of the \rpackage{glmnet} (proposed) against the \rpackage{glmnet} (default) and LARS.

We used an image from the MNIST dataset \citep{lecun1998gradient_MNIST} for the experiment, resizing it to $32\times 32$ pixels.
First, the image was compressed as follows.
The image matrix was vectorized in column-major order to form $\bm{\theta} \in \mathbb{R}^{1024}$.
Let $N'$ denote the dimension of the compressed data.
We generated a random projection matrix $\bm Z \in \mathbb{R}^{N' \times 1024}$, where each element $Z_{ij}$ was drawn independently from $\mathcal{N}(0,1)$.
The vector $\bm{\theta}$ was then compressed into $\bm{y} = \bm Z\bm{\theta} \in \mathbb{R}^{N'}$.
In this experiment, we set the compressed dimension to $N' = 700$.
This process reduces the dimensionality from 1024 to $N'$, effectively compressing the data.

Subsequently, we reconstructed the original image $\bm{\theta}$ using the compressed vector $\bm{y}$ and the projection matrix $\bm Z$.
By employing a two-level wavelet basis matrix $\bm \Psi$, the reconstruction corresponds to solving the following lasso problem:
\begin{equation*}
    \hat{\bm{\beta}}=
    \underset{\boldsymbol{\beta}}{\text{argmin}}\
    \frac{1}{2N'}
    \| \boldsymbol{y}-\bm{X}\boldsymbol{\beta}\|_2 ^2 
    +\lambda \| \boldsymbol{\beta} \|_1, 
\end{equation*}
where $\bm{X}=\bm Z\bm \Psi \in \mathbb{R}^{N' \times 1024}$. 
The reconstructed image is obtained by $\hat{\bm{\theta}}=\bm \Psi \hat{\bm{\beta}}$.
Using this formulation, we evaluated the performance of the proposed method.

\begin{figure}[t]
    \centering
    \begin{subfigure}[b]{0.32\textwidth}
        \centering
        \includegraphics[width=\textwidth]{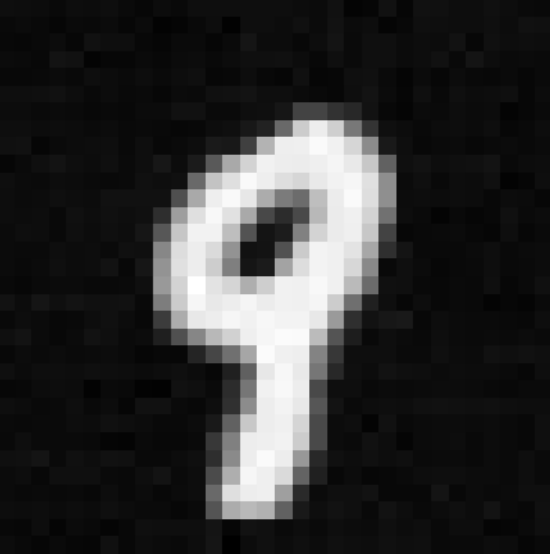}
        \caption{Original image}
        \label{fig:original_image}
    \end{subfigure}
    
    \vspace{1em}
    
    \begin{subfigure}[b]{0.32\textwidth}
        \centering
        \includegraphics[width=\textwidth]{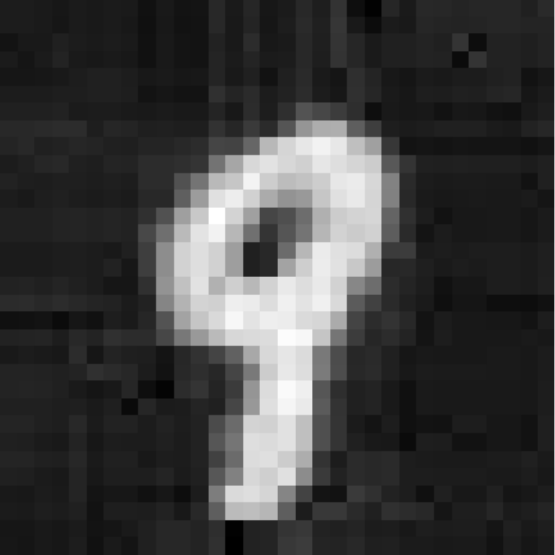}
        \caption{\rpackage{glmnet} (default)}
        \label{fig:reconst_def}
    \end{subfigure}
    \hfill
    \begin{subfigure}[b]{0.32\textwidth}
        \centering
        \includegraphics[width=\textwidth]{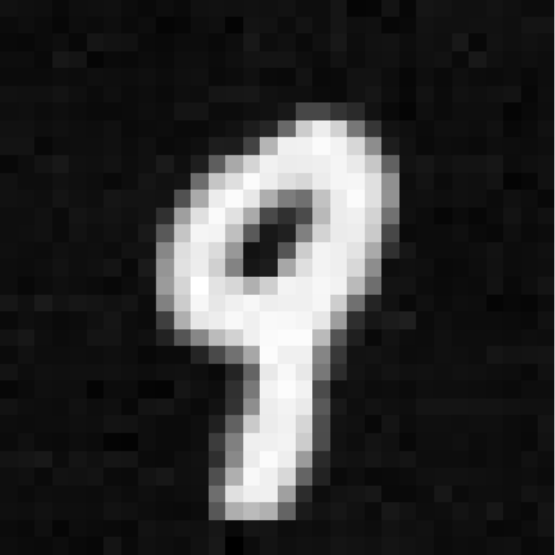}
        \caption{\rpackage{glmnet} (proposed)} 
        \label{fig:reconst_tuned}
    \end{subfigure}
    \hfill 
    \begin{subfigure}[b]{0.32\textwidth}
        \centering
        \includegraphics[width=\textwidth]{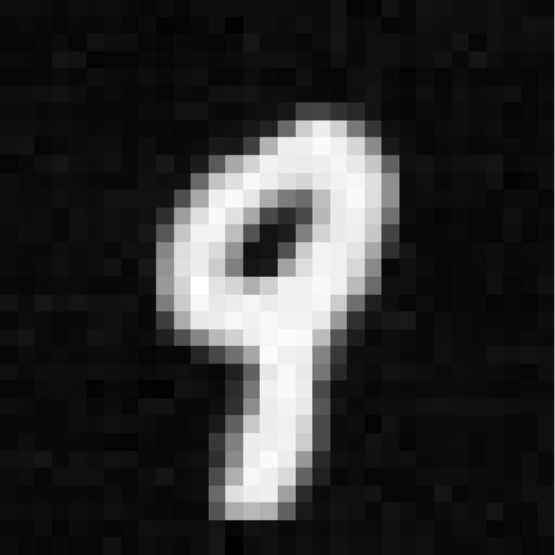}
        \caption{LARS}
        \label{fig:reconst_lars}
    \end{subfigure}

    \caption{Visual comparison of reconstruction results. 
    (a) Original image. 
    The bottom row displays the reconstructed images along with their RMSE values calculated against the original image (on a $[0, 255]$ scale);
    (b) default \rpackage{glmnet} (RMSE: $31.57$);
    (c) \rpackage{glmnet} tuned by the proposed method (RMSE: $14.53$); and 
    (d) LARS (RMSE: $12.22$).
    Notably, regarding computation time, the \rpackage{glmnet} (proposed) required only $13.2~\mathrm{s}$, whereas LARS required $53.4~\mathrm{s}$.
    This demonstrates that our approach successfully identifies an appropriate configuration that is both accurate and computationally efficient.}
    \label{fig:comparison_pyramid}
\end{figure}

Figure \ref{fig:comparison_pyramid} illustrates the reconstruction results.
To quantify the reconstruction quality, we evaluated RMSE between the reconstructed and original images on the pixel value scale $[0, 255]$.
The \rpackage{glmnet} (default) yielded a high RMSE of $31.57$, resulting in a degraded image with reduced sharpness.
In contrast, the proposed method achieved an RMSE of $14.53$, which is significantly lower than the default and comparable to the RMSE of $12.22$ obtained by the exact solution of LARS.
Regarding computational efficiency, the \rpackage{glmnet} (proposed) required only approximately one-fourth of the computation time of LARS.
These results demonstrate that our proposed framework successfully tunes the configuration to achieve accuracy comparable to LARS while maintaining significantly lower computational cost.

\subsection{Discussion}\label{sec:discussion}
Our results demonstrate that the proposed method successfully tunes the configuration to achieve accuracy comparable to that of the exact solution of LARS, while approximately satisfying the specified computation time constraint, $T_{\text{hope}}$.
This suggests that our framework successfully tunes an appropriate configuration that overcomes the limitations inherent in the default configuration.
The observed improvement in test error is primarily attributable to the expanded search range for $\lambda$.
This wider range enables cross-validation to identify optimal $\lambda$ values that restrictive default grids often miss.
Regarding computation time, the prediction accuracy of \texttt{glmnet-MLP} proved reasonably reliable.
This accuracy enabled the selection of a configuration that adhered to the time constraint, $T_{\text{hope}}$.

However, a primary limitation of our framework is the size of the training dataset.
Our model was trained using a synthetic dataset generated from multivariate normal distributions within specific ranges of sample size $n$ and dimension $p$.
As it is practically unachievable to learn the characteristics of all possible data distributions, this dependency on the training dataset is unavoidable.
In particular, caution is required when extrapolating to cases where $N$ or $p$ exceeds the upper bounds of the aforementioned training range.
Nevertheless, the compressed sensing experiment provided a promising indication of robustness.
In this case, although $N$ and $p$ were within the training range, the structural properties of the design matrix $\bm{X}$ differed from the multivariate normal assumption used in training.
The successful application in this context demonstrates the potential applicability of our method to datasets with design matrices outside the training distribution.
Finally, regarding hardware dependency, computation time varies across different computing environments.
However, from a practical standpoint, the order of magnitude is often more critical than precise timing.
Minor deviations in seconds are generally acceptable in real-world applications, provided the algorithm operates within the expected time scale.

\section{Conclusion}
In this study, we established a data-driven framework for configuration tuning of \rpackage{glmnet} by learning from large-scale artificial datasets.
Our approach explicitly models the tradeoff between accuracy and computation time.
This capability enables the identification of a configuration that achieves accuracy comparable to LARS while satisfying user-specified time constraints.

In future work, we aim to address the limitations discussed in Section~\ref{sec:discussion}.
Specifically, we plan to enhance the generalizability of the model by expanding the training dataset to include a wider range of sample sizes and dimensions ($N, p$) and diverse data distributions.
Furthermore, extending this framework to other families of generalized linear models (GLMs) supported by \texttt{glmnet} (e.g., logistic and Poisson regression) represents a promising avenue, given their shared algorithmic structure.

\clearpage
\begin{flushleft}
    \LARGE \textbf{Appendix}  
\end{flushleft}
\appendix

\section{Details of the summary dataset}\label{appendix:parameter}

\subsection{Hyperparameters for summary dataset}
In Section \ref{datagene}, we described the generation of the design matrix $\bm{X}$,$ \bm y$ to obtain the summary dataset for training the \rpackage{glmnet} MLP.
Then, we need to specify the parameters $\Sigma, \bm{\beta}, \sigma$ to generate $\bm{X}, \bm y$.
This appendix provides the specific details of these parameter settings.

\paragraph{Structure of the true covariance matrix $\Sigma$.}
We employed four types of covariance matrix for $\bm{\Sigma}$:
\begin{enumerate}
    \item \textbf{Compound symmetry covariance matrix:}
    \[ \bm{\Sigma}=(1-\rho) \ \bm{I}_p + \rho \,\bm{1}_p\bm{1}_p^\top=
    \begin{pmatrix}
        1      & \rho     & \dots  & \rho\\
        \rho     & 1      & \dots  & \rho\\
        \vdots & \vdots & \ddots & \vdots \\
        \rho     & \rho     & \dots  & 1
    \end{pmatrix}\quad (0<\rho<1)
    \]

    \item \textbf{AR(1) covariance matrix:}
    \[
    \begin{pmatrix}
        1      & \rho     & \dots  & \rho^{p-1} \\
        \rho     & 1      & \dots  & \rho^{p-2}\\
        \vdots & \vdots & \ddots & \vdots \\
        \rho^{p-1}    & \rho^{p-2}     & \dots  & 1
    \end{pmatrix}\quad (0<\rho<1)
    \]

    \item \textbf{Random structured covariance matrix:}
    The construction procedure for the random structured covariance matrix is based on \citet{HIROSE2017172}.
    The specific steps are as follows:
    \begin{enumerate}
        \item Define the set $\mathcal{D}=[-0.75,-0.25]\cup[0.25,0.75]$, and construct a $p \times p$ matrix $E$, where
        each element $E_{i,j}$ is drawn independently from a uniform distribution $U(\mathcal{D})$.
        \item Assign $0$ to some off-diagonal elements of the matrix $E$ generated.
        The number and specific locations for these assignments are determined by a uniform random selection.
        \item Compute $\tilde{E}=(E+E^\top)/2$.

        \item Calculate $\tilde{\Omega}=\tilde{E}+(0.1-\lambda_{\min})\mathbf{I}$, where $\lambda_{\min}$ is the minimum eigenvalue of $\tilde{E}$.

        \item Let $L=\text{diag}(\tilde{\Omega}^{-1})$, and then compute $\Omega=L^{\frac{1}{2}}\tilde{\Omega}L^{\frac{1}{2}}$.

        \item Finally, the random structured covariance matrix $C$ is expressed by $C=\Omega^{-1}$.
    \end{enumerate}
    \item \textbf{Inverse of the random structured covariance matrix:}
    We adopt the inverse of $C$ as the covariance matrix.
\end{enumerate}

\paragraph{Structure of the true coefficient $\bm{\beta}$.}
We prepared the following four structural patterns for $\bm{\beta} \in \mathbf{R}^p$:

\begin{enumerate}
    \item $\lfloor \frac{p}{2}  \rfloor$ elements are 1, and the others are 0.
    \item $\lfloor \frac{p}{10}  \rfloor$ elements are 1, and the others are 0.
    \item $\lfloor \frac{p}{2}  \rfloor$ elements are generated from $\mathcal{N}(0,1)$, and the others are 0.
    \item $\lfloor \frac{p}{10} \rfloor$ elements are generated from $\mathcal{N}(0,1)$, and the others are 0.
\end{enumerate}
In all cases, the positions of the $\bm \beta$ elements are randomly permuted.

\paragraph{Structure of the true error standard deviation $\sigma$.}
We employed two settings for the noise level $\sigma \in \mathbf{R}$:

\begin{enumerate}
    \item $\sigma = 1$
    \item $\sigma = \frac{p}{10}$
\end{enumerate}

\subsection{Scope of the summary dataset}\label{detail_summarydataset}
\begin{figure}[t]
    \centering
    \includegraphics[width=\linewidth]{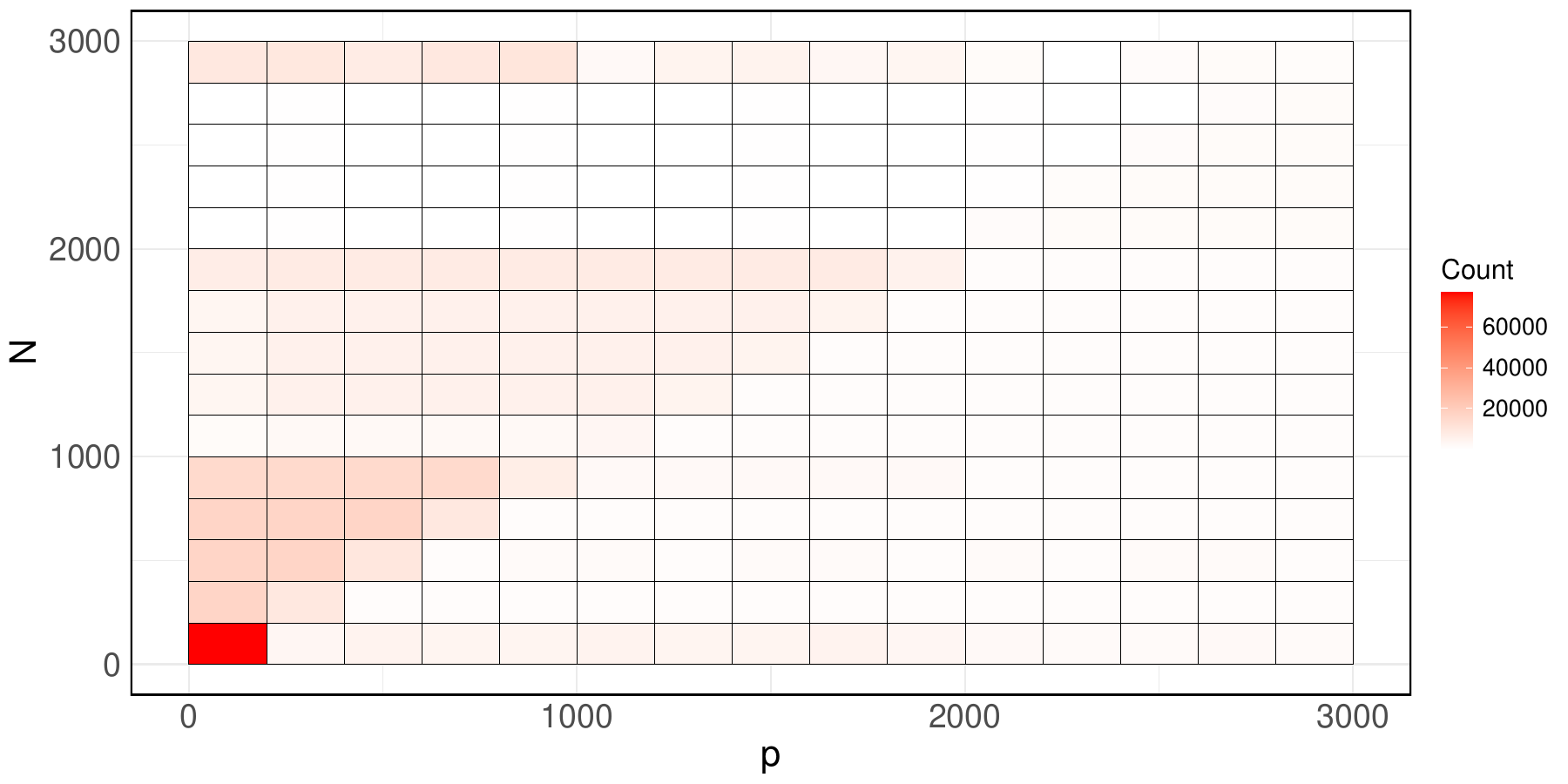}
    \caption{Heatmap showing the distribution of combinations of \(N\) and \(p\) in the summary dataset.
    The color intensity represents the number of data points; redder regions indicate a higher concentration of samples.
    White regions indicate sparse sampling (low count), not necessarily the absence of dataset.}
    \label{heatmap}
\end{figure}

This section describes the scope of the summary dataset.
Specifically, the distribution of sample size $N$ and the number of predictors $p$ directly determines 
the applicable range of our proposed method.
Figure~\ref{heatmap} illustrates the distribution of $N$ and $p$ within the dataset.
The dataset spans a broad range of dimensions, explicitly defined by the ranges $N \in [30, 3000]$ and $p \in [10, 2980]$.
To ensure the accuracy of computation time measurements, we avoided large-scale parallelization.
This constraint significantly increased the total time required to generate the summary dataset.
Consequently, we employed a denser sampling strategy in regions where $N$ and $p$ are small.

\subsection{Computational environment}
All numerical experiments were conducted on a server running Ubuntu 24.04.1 LTS (Linux kernel 6.8.0), equipped with an AMD EPYC 7763 64-Core Processor (up to 3.5 GHz) and 2~TB of DDR4-3200 ECC memory. 
Computational tasks were implemented in \texttt{R} version 4.3.3. 
Parallel processing with 10 logical cores was employed during the generation of summary datasets to enhance computational efficiency, using the \rpackage{doParallel} (v1.0.17) and \rpackage{foreach} (v1.5.2) packages.
In contrast, other simulation procedures were executed in a single-threaded manner. 
The \texttt{R} environment was linked against the reference BLAS (v3.12.0) and LAPACK (v3.12.0) libraries. 
The versions of \rpackage{glmnet} and LARS were 4.1.8 and 1.3, respectively.

\section{Details of training strategy and hyperparameters}\label{app:training_details}

In this section, we provide detailed specifications of the training process and the resulting model architecture for \texttt{glmnet-MLP}, described in Section \ref{training_strategy}.

\paragraph{Data preparation.}
The summary dataset was split into 80\%, 10\%, and 10\% for training, validation, and testing, respectively.
As mentioned in the main text, target variables were log-transformed and standardized.

\paragraph{Optimization setup.}
The Bayesian optimization was performed using the \texttt{BoTorchSampler} \citep{balandat2020botorch} within \texttt{Optuna}, based on Gaussian process regression and the Expected Improvement acquisition function.
We executed the optimization for 500 trials.
Throughout the process, the activation function was fixed to the Swish function \citep{ramachandran2017searching}:
\begin{equation*}
    f(x)=\frac{x}{1+e^{-x}}.
\end{equation*}
The number of epochs was set to 500, and the minibatch size to 20,263.
The search space for the optimization was defined as follows:
\begin{itemize}
    \item Number of hidden layers: $\{1, 2, 3\}$;
    \item Number of units per layer: $\{1, 2, \dots, 64\}$;
    \item Learning rate: $[10^{-5}, 10^{-1}]$ (log scale).
\end{itemize}

\paragraph{Resulting model hyperparameters.}
The optimization resulted in selecting a three-layer network, with 64, 61, and 57 units in the respective hidden layers.
The optimal learning rate was approximately $7.6 \times 10^{-4}$.
This configuration was adopted as the final \texttt{glmnet-MLP} for evaluation.

\section{\texttt{R} Package \texttt{glmnetconf}}\label{glmnetconf}
We developed the \texttt{R} package \rpackage{glmnetconf} to implement our tuning method and ensure its accessibility to a wide audience.
Our implementation includes not only a configuration tuning method but also a method for selecting the $\tt R$ package.
As \rpackage{lars} and \rpackage{glmnet} each possess distinct advantages,
the appropriate choice depends on the objective of the user.
Specifically, if an exact solution is required without considering computation time, \rpackage{lars} is the optimal choice.
Conversely, if computational efficiency is prioritized, \rpackage{glmnet} is preferable.
Therefore, we implemented the function to select the $\tt R$ package based on the dataset $\bm{X}, \bm y$ and $T_\text{hope}$ in our package.

\subsection{Workflow of \rpackage{glmnetconf}}
\begin{figure}[t]  
    \centering
    \includegraphics[width=1\textwidth]{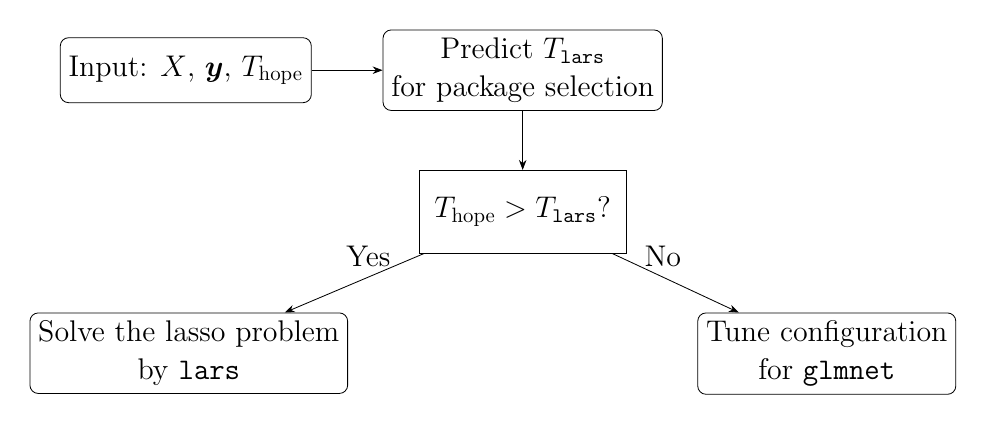}  
    \caption{Workflow of our package \rpackage{glmnetconf}}  
    \label{fig:workflow}  
\end{figure}

We assumed that we have a dataset $\bm{X}, \bm y$ and desired computation time $T_{\text{hope}}$.
Figure \ref{fig:workflow} shows the workflow of our proposed package.
First, our framework determines which package to employ.
We predict the computation time of \rpackage{lars}, denoted as $T_\mathtt{lars}$, based on $\bm{X}, \bm y$.
If the predicted $T_\mathtt{lars}$ is smaller than $T_\text{hope}$, our framework selects \rpackage{lars} to ensure an exact solution.
By contrast, if $T_\mathtt{lars}$ exceeds $T_\text{hope}$, our framework selects \rpackage{glmnet}. 
In this scenario, the configuration for \rpackage{glmnet} is tuned by our proposed method described in Section \ref{sec:proposed_method}.

\subsection{Prediction model for the \rpackage{lars} computation time}\label{larsMLP}
Similar to the \texttt{glmnet-MLP}, we constructed a predictive model to forecast the computation time of \rpackage{lars}, denoted as $T_{\rpackage{lars}}$.
The input features consist of the sample size $N$, dimension $p$, and selected eigenvalues of the sample covariance matrix of $\bm{X}$ ($\gamma_{\pm 1}, \dots, \gamma_{\pm 5}$).
The output is the predicted computation time $T_{\rpackage{lars}}$.

The training dataset for the \texttt{lars-MLP} was collected during the generation of the summary dataset described in Section \ref{datagene}.
The resulting dataset comprises $68,013$ samples.
Using this dataset, we trained the model employing the Adam optimizer.

The network architecture and hyperparameters were determined via Bayesian optimization, following the same protocol and search space as the \texttt{glmnet-MLP}.
The optimization yielded a three-layer hidden network with 45, 44, and 37 units in the respective layers, and a learning rate of approximately $1.08 \times 10^{-3}$.
Consistent with the \texttt{glmnet-MLP}, we employed the Swish activation function and set the number of epochs to 500.
However, the batch size was set to 1700 for this model.

\begin{figure}[t]
    \centering
    \includegraphics[width=\linewidth]{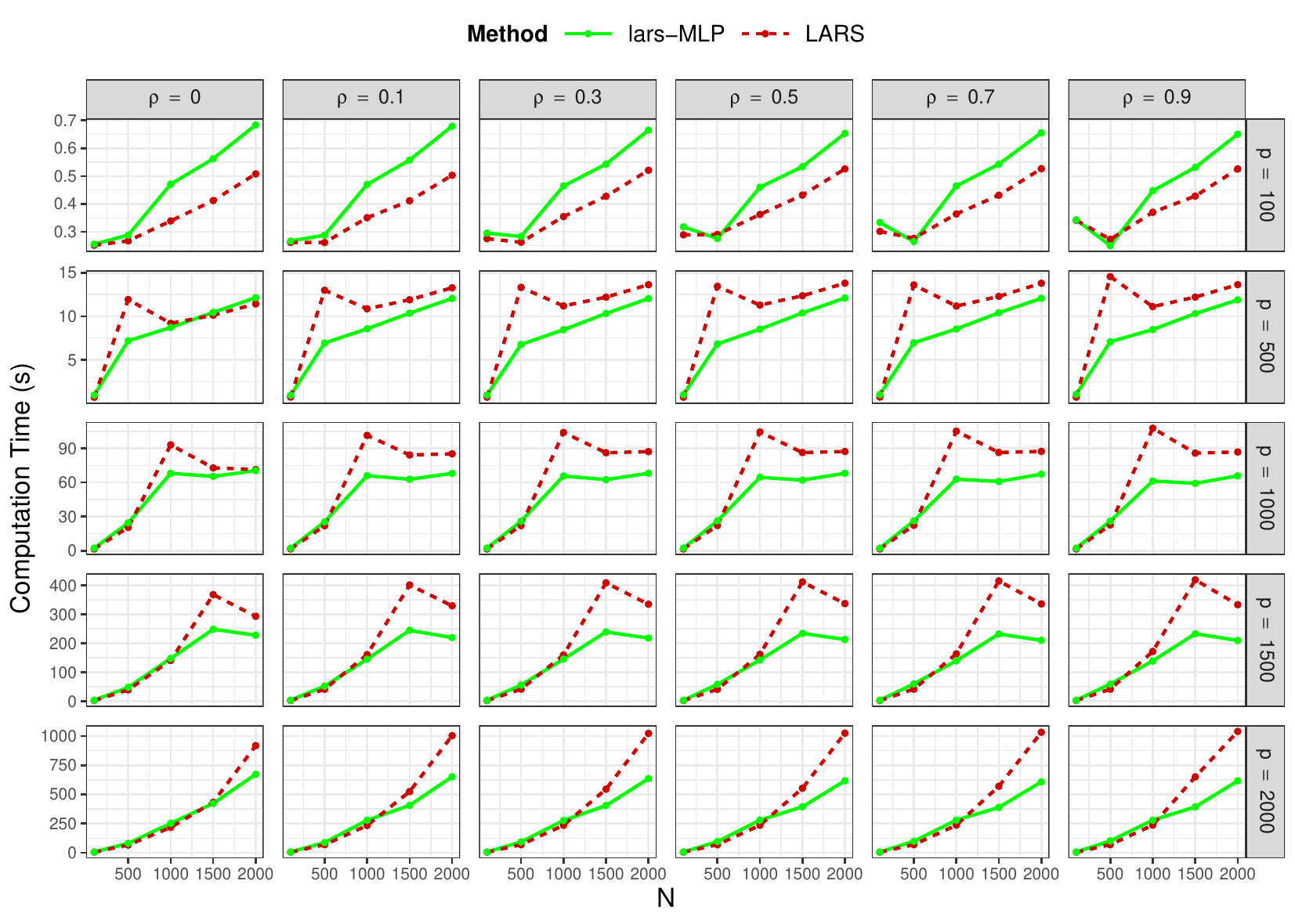}
    \caption{Comparison of predicted versus actual computation times for \rpackage{lars} across different sample sizes $N$.
    The layout is identical to that of Figure~\ref{fig:simulation_rmse}.
    The \texttt{lars-MLP} effectively captures the overall trend of the computation time.
    However, it underestimates the actual computation time when both $N$ and $p$ are large.}
    \label{fig:simulation_time_prediction}
\end{figure}

Figure~\ref{fig:simulation_time_prediction} compares the predicted computation time with the actual runtime of \rpackage{lars}, using the simulation dataset described in Section~\ref{sec: numericalexp}.
The model effectively captures the overall trend of the computation time.
However, it underestimates the runtime when both $N$ and $p$ are large.
This bias is likely owing to the scarcity of training samples in high-dimensional regions, constrained by the high computational cost of data generation.

\subsection{Usage example}
This section demonstrates the usage of the \rpackage{glmnetconf} package.
The primary function, \texttt{auto\_lasso()},
automates the entire tuning process.
Specifically, it automatically selects the appropriate package and tunes the configuration 
based on the input dataset and $T_\text{hope}$.

Listing \ref{lst:glmnetconf_sample} shows a usage example with the synthetic dataset, where $T_\text{hope}=20$ s.
The dataset is generated via the function \texttt{data\_generation}, following the same simulation settings described in Section \ref{sec: numericalexp}.
In this example, we set the sample size to $N=1500$, dimension to $p=800$, and correlation to $\rho=0.5$.
The script not only executes the proposed automated workflow via \texttt{auto\_lasso()} but also compares its predictive performance against the standard usage of \texttt{cv.glmnet} (default configuration).
This comparison illustrates how the proposed method achieves competitive accuracy while satisfying the time constraint.

\medskip

\begin{minipage}{\linewidth}
    \begin{lstlisting}[caption={Usage example of \rpackage{glmnetconf}. The script demonstrates the proposed tuning method with a time constraint and compares its test error against the default configuration of \rpackage{glmnet}.}, label={lst:glmnetconf_sample}]
library(glmnetconf); library(glmnet)

# Setup: Generate X_train, y_train, X_test, y_test.
dat <- data_generation(N_train = 1500, N_test = 100, p = 800, rho = 0.5, sparse_rate = 0.5, sigma = 1)
X_train <- dat$X_train
y_train <- dat$y_train
X_test  <- dat$X_test
y_test  <- dat$y_test

# --- 1. Proposed Method (glmnetconf) ---
# Perform automatic tuning with a time constraint T_hope = 20s
fit_glmnetconf <- auto_lasso(X_train, y_train, new_x = X_test, T_hope=20)
mse_glmnetconf <- mean((y_test - fit_glmnetconf$prediction)^2)

# --- 2. Benchmark (glmnet default) ---
# Standard cross-validation with default settings
fit_glm  <- cv.glmnet(X_train, y_train, alpha = 1)
pred_glm <- predict(fit_glm, newx = X_test, s = "lambda.min")
mse_glm  <- mean((y_test - pred_glm)^2)

# --- 3. Performance Comparison ---
print(c(glmnetconf = mse_glmnetconf, glmnet = mse_glm))

# (Optional) Check tuned configuration and Pareto front
# print(fit_glmnetconf$configuration)
# print(fit_glmnetconf$Pareto_front)
\end{lstlisting}
\end{minipage}

\section*{Acknowledgements}
We would like to thank Editage (www.editage.jp) for English language editing.



\begin{thebibliography}{}

\bibitem[Akiba et~al., 2019]{akiba2019optuna}
Akiba, T., Sano, S., Yanase, T., Ohta, T., and Koyama, M. (2019).
\newblock Optuna: A next-generation hyperparameter optimization framework.
\newblock In {\em Proceedings of the 25th ACM SIGKDD International Conference on Knowledge Discovery \& Data Mining}, KDD '19, page 2623–2631, New York, NY, USA. Association for Computing Machinery.

\bibitem[Balandat et~al., 2020]{balandat2020botorch}
Balandat, M., Karrer, B., Jiang, D.~R., Daulton, S., Letham, B., Wilson, A.~G., and Bakshy, E. (2020).
\newblock {BoTorch: A Framework for Efficient Monte-Carlo Bayesian Optimization}.
\newblock In {\em Advances in Neural Information Processing Systems 33}.

\bibitem[Beck and Teboulle, 2009]{amir2009fista}
Beck, A. and Teboulle, M. (2009).
\newblock A fast iterative shrinkage-thresholding algorithm for linear inverse problems.
\newblock {\em SIAM Journal on Imaging Sciences}, 2(1):183--202.

\bibitem[B{\o}velstad et~al., 2007]{bocelstad2007maicroarraylasso}
B{\o}velstad, H., Nyg{\aa}rd, S., St{\o}rvold, H., Aldrin, M., Borgan, {\O}., Frigessi, A., and Lingj{\ae}rde, O. (2007).
\newblock Predicting survival from microarray data—a comparative study.
\newblock {\em Bioinformatics}, 23(16):2080--2087.

\bibitem[Boyd et~al., 2011]{boyd2011admm}
Boyd, S., Parikh, N., Chu, E., Peleato, B., and Eckstein, J. (2011).
\newblock Distributed optimization and statistical learning via the alternating direction method of multipliers.
\newblock {\em Foundations and Trends{\textregistered} in Machine learning}, 3(1):1--122.

\bibitem[Cand{\`e}s and Wakin, 2008]{candes2008introduction}
Cand{\`e}s, E.~J. and Wakin, M.~B. (2008).
\newblock An introduction to compressive sampling.
\newblock {\em IEEE signal processing magazine}, 25(2):21--30.

\bibitem[Cs{\'a}rdi, 2019]{cranlogs2019}
Cs{\'a}rdi, G. (2019).
\newblock {\em cranlogs: Download Logs from the 'RStudio' 'CRAN' Mirror}.
\newblock R package version 2.1.1.

\bibitem[Daubechies et~al., 2004]{daubechies2004ista}
Daubechies, I., Defrise, M., and De~Mol, C. (2004).
\newblock An iterative thresholding algorithm for linear inverse problems with a sparsity constraint.
\newblock {\em Comm. Pure Appl. Math.}, 57(11):1413--1457.

\bibitem[Efron et~al., 2004]{efronleast2004}
Efron, B., Hastie, T., Johnstone, I., and Tibshirani, R. (2004).
\newblock Least angle regression.
\newblock {\em The Annals of Statistics}, 32(2):407--499.

\bibitem[Friedman et~al., 2010]{friedmanregularization2010}
Friedman, J.~H., Hastie, T., and Tibshirani, R. (2010).
\newblock Regularization {Paths} for {Generalized} {Linear} {Models} via {Coordinate} {Descent}.
\newblock {\em Journal of Statistical Software}, 33:1--22.

\bibitem[Fu, 1998]{Fu1998shooting}
Fu, W.~J. (1998).
\newblock Penalized regressions: The bridge versus the lasso.
\newblock {\em Journal of Computational and Graphical Statistics}, 7(3):397--416.

\bibitem[Hebiri and Lederer, 2013]{hebiri2013corlasso}
Hebiri, M. and Lederer, J. (2013).
\newblock How correlations influence lasso prediction.
\newblock {\em IEEE Trans. Inf. Theor.}, 59(3):1846–1854.

\bibitem[Hirose et~al., 2017]{HIROSE2017172}
Hirose, K., Fujisawa, H., and Sese, J. (2017).
\newblock Robust sparse gaussian graphical modeling.
\newblock {\em Journal of Multivariate Analysis}, 161:172--190.

\bibitem[LeCun et~al., 1998]{lecun1998gradient_MNIST}
LeCun, Y., Bottou, L., Bengio, Y., and Haffner, P. (1998).
\newblock Gradient-based learning applied to document recognition.
\newblock {\em Proceedings of the IEEE}, 86(11):2278--2324.

\bibitem[Lu and Li, 2015]{Lu2015astrolasso}
Lu, Y. and Li, X. (2015).
\newblock Estimating stellar atmospheric parameters based on lasso and support-vector regression.
\newblock {\em Monthly Notices of the Royal Astronomical Society}, 452(2):1394--1401.

\bibitem[Massias et~al., 2018]{massias_2018_celer}
Massias, M., Gramfort, A., and Salmon, J. (2018).
\newblock Celer: a fast solver for the lasso with dual extrapolation.
\newblock In Dy, J. and Krause, A., editors, {\em Proceedings of the 35th International Conference on Machine Learning}, volume~80 of {\em Proceedings of Machine Learning Research}, pages 3315--3324. PMLR.

\bibitem[Osborne et~al., 2000]{Osborne2000homotopy}
Osborne, M., Presnell, B., and Turlach, B. (2000).
\newblock A new approach to variable selection in least squares problems.
\newblock {\em IMA Journal of Numerical Analysis}, 20(3):389--403.

\bibitem[Ramachandran et~al., 2017]{ramachandran2017searching}
Ramachandran, P., Zoph, B., and Le, Q.~V. (2017).
\newblock Searching for activation functions.
\newblock {\em arXiv preprint arXiv:1710.05941}.

\bibitem[Rumelhart et~al., 1986]{rumelhart_learning_1986}
Rumelhart, D.~E., Hinton, G.~E., and Williams, R.~J. (1986).
\newblock Learning representations by back-propagating errors.
\newblock {\em Nature}, 323(6088):533--536.

\bibitem[Tibshirani, 1996]{tibshiraniregression1996}
Tibshirani, R. (1996).
\newblock Regression {Shrinkage} and {Selection} {Via} the {Lasso}.
\newblock {\em Journal of the Royal Statistical Society: Series B (Methodological)}, 58(1):267--288.

\end{thebibliography}

\end{document}